# Mechanisms and Controversies of tACS (Transcranial Alternating Current Stimulation)

Julien Modolo*[1], Mengsen Zhang*[2,3], Joan Duprez[1], Flavio Frohlich[3,4,5,6,7]

1: Univ Rennes, LTSI - U1099, F-35000 Rennes, France
2: Department of Psychiatry, University of North Carolina at Chapel Hill, Chapel Hill, NC, 27599, United States
3: Carolina Center for Neurostimulation, University of North Carolina at Chapel Hill, Chapel Hill, NC, 27599, United States
4: Department of Neurology, University of North Carolina at Chapel Hill, Chapel Hill, NC, 27599, United States
5: Department of Biomedical Engineering, University of North Carolina at Chapel Hill, Chapel Hill, NC, 27599, United States
6: Department of Cell Biology and Physiology, University of North Carolina at Chapel Hill, Chapel Hill, NC, 27599, United States
7: Neuroscience Center, University of North Carolina at Chapel Hill, Chapel Hill, NC, 27599, United States

*: co-first authors.
**Corresponding author:** julien.modolo@inserm.fr

**Abstract**. The use of small amplitude electric currents delivered through the scalp, termed transcranial current stimulation (tCS, tDCS / tACS when using DC / AC currents) holds considerable promise for developing safe and effective treatments for central nervous system disorders. Initially welcomed with skepticism due to significant gaps of knowledge in terms of neurophysiology and biophysical mechanisms, tCS is maturing as a technology while its mechanisms of action are gradually being elucidated. However, there remain open questions about the mechanisms of action that warrant clarification to bring tCS to its full potential. In this review focused on tACS, we make an attempt at providing an overview of the converging experimental evidence, from results obtained in various species, regarding the mechanisms of action. We also highlight the remaining points of uncertainty regarding potential confounds, and propose possible experimentally testable solutions to address those issues. Finally, we outline how a continued focus on deepening our understanding of tACS mechanisms might provide significant insights into fundamental, long-standing questions in neuroscience.

**Introduction**

Transcranial current stimulation (tCS) encompasses the use of scalp electrodes that apply a low-intensity (typically up to 2 mA) to the scalp with the objective to impact the activity of the underlying brain tissue. While the use of non-invasive electrical stimulation technique is certainly not new (its origins can be traced to the Ancient Egypt, (Sarmiento et al., 2016)), its impact on the neuroscience research and clinical communities took off more recently (early 2000's, with seminal works such as those by Paulus and Nietsche (Nitsche & Paulus, 2000)). While tCS was initially developed using a constant current (DC stimulation, hence the name tDCS), novel modalities have emerged, including the popular tACS (AC stimulation, using a sinusoidal current at a given frequency) and ones with more complex waveforms such as tRNS (transcranial random noise stimulation, using a noisy input current). In this review, we will focus mostly on tACS, which offers a wide range of potential applications due to its capacity to adapt to the frequency of endogenous activity in targeted brain circuits.

The principle of tCS is simple and appealing: a low-cost, user-friendly technology that modulates brain function and behavior, with side effects that are either absent or very minor (e.g., itching or burning sensation below the stimulating electrode). During the "re-discovery" of tCS in the 1950's (for an exhaustive historical review, see (Sarmiento et al., 2016)), the effects that were reported were behavioral such as increased alertness, and neurophysiological such as a modulation of cortical excitability, pointing to a modulation of cortical circuits by the applied stimulation. Within the last 15 years, tCS has been shown, in different contexts and in healthy volunteers and patients with a wide range of neurological and psychiatric disorders,.

However, as the saying goes, "*If it sounds too good to be true, it probably is*", and it is for good reasons that tCS has been historically facing some skepticism from the research community. Any neuroscientist faced for the first time with the general principles of tCS will invariably ask the same questions: how can there be *any* effect on brain tissue activity given the low magnitude of the induced electric field (1-2 V/m at most)? How can membrane depolarizations smaller than 1 mV could impact neuronal dynamics, when the firing threshold is approximately 15 to 20 mV higher than the resting potential? Why do reported effects take time to occur (typically, at least 5 to 10 minutes is the "golden duration" used in most studies that is required to reach detectable effects, deriving from (Nitsche & Paulus, 2001) and (Nitsche et al., 2003), among others)? All of those questions are legitimate ones, and answering them might the key to unlock the true potential of tACS, as detailed below.

A major challenge in understanding the effects of tACS is that a considerable portion of the literature on the effects of electric fields on neuronal activity uses 1) supra-threshold electric fields, i.e. triggering action potentials; 2) square biphasic pulses, involving a large duty cycle; 3) stimulation delivered intracranially (electrodes placed within the tissue). However, tACS differs in all these dimensions: 1) it induces subthreshold electric fields (on the order of 0.2-2 V/m *in situ*); 2) tACS applies sinusoidal electrical currents; and 3) stimulation is delivered non-invasively by electrodes placed on the scalp, with a small fraction of the current effectively reaching the cortex (Asamoah et al., 2019), after crossing several tissues with different electrical properties. Therefore, when examining results from the tACS literature, those have to be taken with caution, by carefully accounting for the experimental conditions and hardware that were used, to ensure that results are indeed applicable to the case of tACS.

In this review, we will first discuss the best-established effects of tACS on central nervous activity. Second, we will attempt at exploring the main gaps in understanding the mechanisms of action of tACS, which will lead to identify several controversies that have emerged over the last few years, including the possibility of (at least in part) contribution from peripheral nerve activation in the reported effects of tACS. Finally, we will suggest several possible directions to move tACS research forward and address those gaps, as an effort to bring this field closer to reproducible, reliable therapies in a wide spectrum of neurological disorders.

**tCS documented effects and mechanisms**

*Established mechanisms of tCS*

It was established over 15 years ago that weak (albeit DC) electric fields (on the order of those induced by tACS) could impact the phase of neuronal spiking and associated spike timing *in vitro* (T. Radman et al., 2007), providing a first evidence in the mechanistic chain of events linking the *in situ* electric field and reported neurophysiological effects. From the mechanistic point of view, the origin of this effect can be traced back to the differential equations that govern membrane dynamics (e.g., Hodgkin-Huxley type equations): depending on the current state of dynamical variables (membrane potential and ionic channel gating variables), the membrane voltage is more sensitive to inputs, and intrinsic dynamics can amplify their impact. This can also be linked with the concept of Phase Response Curves (PRC, (Winfree, 1980), which describe the phase-dependence of a system's response as a function of the phase where the input is received. Crucially, such modulation of spike timing (i.e., neuronal phase) is the very first biological consequence of tACS within the central nervous system, and it is notable that a well-defined mechanism describes it relatively parsimoniously.

More recently, (Johnson et al., 2020) provided an *in vivo* confirmation in monkeys that tACS-induced electric fields on the order of 1 V/m can modulate spike timing in 15-20% of neurons, hence the phase of oscillations with a phase entrainment, i.e. an alignment of endogenous oscillations' phase with the stimulus. Such results contribute to transform our perspective on tACS effects, and challenge the initial view that the induced electric fields were too low to induce spikes, to provide a novel framework in which the phase modulation of endogenous oscillations might be a key mechanism. We can then formulate a few words of caution against our initial intuition towards tACS effects, and that "subthreshold for action potential initiation does not mean subthreshold for biological effects". This very specific point has also significant implications, albeit being outside the scope of this review, for the international guidelines protecting the general public and workers from potentially harmful effects from electromagnetic fields exposure, and that were developed at the time on the premise that the threshold for biological effects should be chosen at the threshold for action potential initiation.

The effect of weak electric field on neuronal activity crucially depends on the endogenous activities of the brain. This is a natural consequence of the fact that a weak electric field (<5 mV/m) is usually insufficient to bring the membrane potential pass the firing threshold of a neuron at rest, such that any observable effects of the field need to be in addition to or in synergy with other activities. The brain is a highly active system at "rest" (Raichle, 2011). Electrical activity of neurons is a major energy consumer of the resting brain (Harris et al., 2012). The intrinsic firing rates of neurons range from below 0.1 Hz to more than 10 Hz follow a log-normal distribution (Buzsáki & Mizuseki, 2014). At the macroscopic level, electric fields in the brain exhibit oscillations at various frequencies from below 1Hz (slow oscillations) to a few hundred Hz (ripples) (Buzsáki & Draguhn, 2004; Buzsáki & György, 2006), which is thought to mainly reflect the synchronization of postsynaptic potentials (Buzsáki et al., 2012). Neuronal spikes are often locked to specific phases of the endogenous oscillations, suggesting that action potentials of neurons are entrained to the macroscopic oscillations (Fröhlich, 2014; Fröhlich & McCormick, 2010; Zhang & Frohlich, 2022).

Interestingly, the endogenous extracellular electric fields in the brain are relatively weak (~1 V/m) and comparable to typical tCS field strength (Fröhlich, 2014; Fröhlich & McCormick, 2010). More importantly, neuronal spiking patterns have been found sensitive to externally applied weak electric fields *in vitro* and *in vivo* (Francis et al., 2003; Fröhlich & McCormick, 2010; Johnson et al., 2020). Intracranial or *in vitro* DC injection in the extracellular space can alter transmembrane potential of neurons and their inter-spike intervals, while intracranial AC injection can entrain neuronal spikes, which become locked to specific phases of the applied field (Fröhlich & McCormick, 2010). These observations are supported by theoretical and computational models (Fröhlich & McCormick, 2010; Thomas Radman et al., 2007). More recently, studies using transcranial stimulations and intracranial recordings in animal models have demonstrated similar entrainment effects (Ali et al., 2013; Huang et al., 2021; Johnson et al., 2020; Ozen et al., 2010). Interestingly, faster spiking interneurons showed a stronger entrainment to both endogenous field and tCS (Huang et al., 2021).

A predominant use of tACS is to entrain and enhance brain oscillations at a target frequency (Neuling et al., 2013; Zaehle et al., 2010). Such enhancement have been shown to affect perception, memory, and various cognitive functions (Herrmann et al., 2013; Marshall et al., 2006; Neuling et al., 2012; Strüber et al., 2014; Thut et al., 2011), and serve as a potential treatment for psychiatric symptoms (Ahn et al., 2019; Alexander et al., 2019; Mellin et al., 2018). From both an engineering and mechanistic perspective, it is key to understand how the amplitude, frequency, and phase of tACS affect the engagement of endogenous oscillations and neuronal activities. Computational modeling and animal studies have shown that the entrainment effect is stronger when the stimulation frequency is closer to that of the endogenous rhythm and when the amplitude of the stimulation is greater (Ali et al., 2013; Huang et al., 2021). The modeling work of Ali et al. (2013) further showed that phase of the stimulation onset affects the relaxation time to full entrainment. These observations closely follow the predictions of dynamical systems theories of coupled oscillators (Kelso, 1995; Strogatz, 2003;

Winfree, 1967). In dynamical system theories, whether or not two oscillators are synchronized (phase-locked) depends on the coupling strength and natural frequency ratio between them. The region in the parameter space (amplitude x frequency) that permits synchronization [ref to figure] is in the shape of an inverted triangle with curved sides, which was known by the work of Vladmir Arnold on circle maps (Arnold, 1965), and commonly referred to as Arnold Tongues. Huang et al. (2021) provided the first empirical observation of an Arnold Tongue in neuronal spike-to-tACS phase-locking in the posterior parietal cortex (PPC) of ferrets. This demonstrates that dynamical systems modeling may play an important role in improving tACS design for better engaging endogenous neuronal activities.

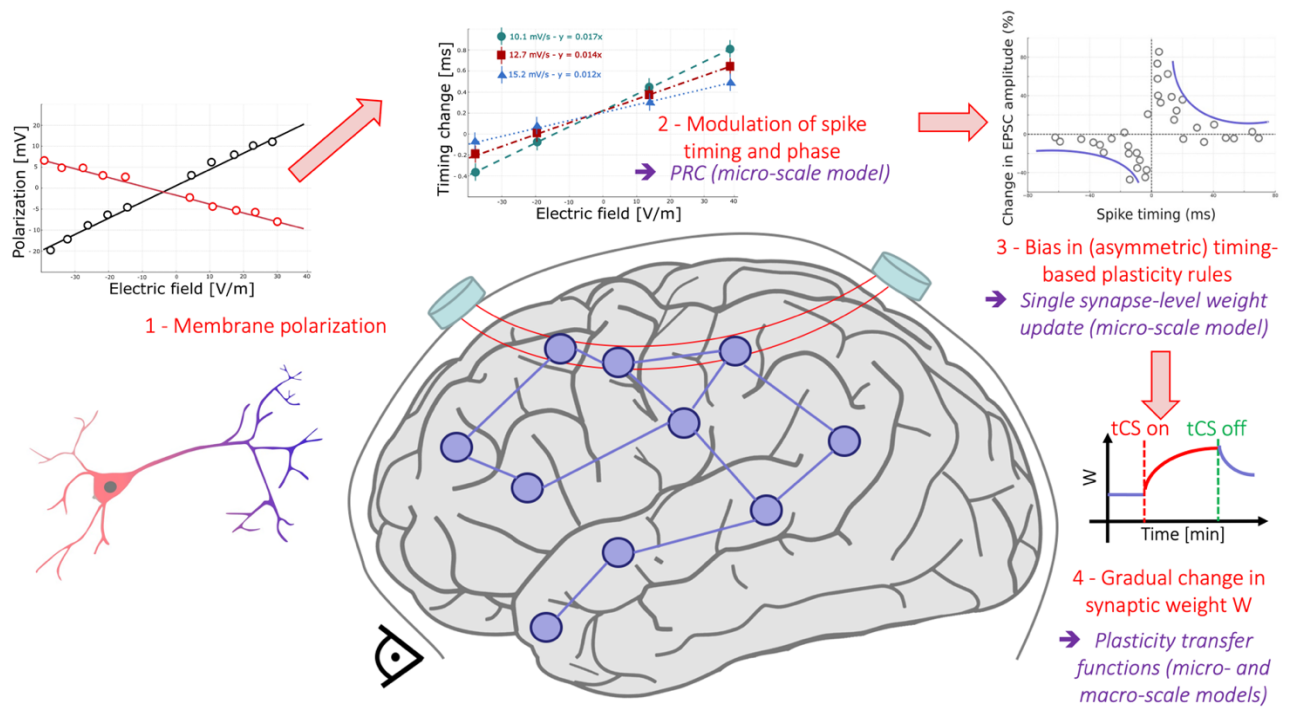

**Figure 1.** Overview of a putative mechanistic chain of tCS mechanisms resulting in neurophysiological and behavioral effects.

**Controversies of tCS: too low, too superficial?**

*Candidate mechanisms of tCS*

One major challenge in understanding tCS effects is that significant variability exists in basically all the parameters of experiments. First and foremost, inter-individual variability in brain morphology can induce *in situ* electric fields that can vary by up to a factor 3 in terms of maximal value (Laakso et al., 2015). Crucially, since the electric fields induced by tCS are low, and likely at the limit of being able to induce biological effects, this variability alone can explain the poor reproducibility that has plagued the field of tCS. However, with a careful design, this is actually possible to overcome this limitation. Second, there is indeed increasing awareness of the need of consistency in tCS experiments, in terms of electrode montage (number of electrodes used to deliver stimulation), stimulation parameters (intensity, frequency, duration), which has led to increased reproducibility such as the replicated improvements in working memory with carefully designed tCS (frequency adapted to the well-known frequency bands involved in this function, i.e. theta and gamma, see (Alekseichuk et al., 2016; Reinhart & Nguyen, 2019)). Therefore, we argue that the lack of reproducible effects that has been present initially, and that was used as an argument that tCS amplitudes were too low to have a meaningful impact on neurophysiology, is gradually being addressed through improved consistency and care in experimental design. Let us mention that recent studies have suggested solutions able to increase the *in situ* electric field magnitude (e.g. using the approach of (Voroslakos et al., 2018)), which should increase further tACS results reproducibility, as also supported by (Zanto et al., 2021).

While acute effects induced by low electric fields on human neurophysiology have been controversial, lasting effects even more so, especially when lasting effects can be reported from minutes to hours after cessation of the stimulation. One of the most immediate explanations to explain such lasting effects would be modulations of neuronal excitability in general (Kasten et al., 2016; Vossen et al., 2015). What are plausible candidates that could explain such neuronal excitability changes after tCS? Those candidates include the modulation of post-synaptic receptors (i.e., receptor trafficking), changes in probability of neurotransmitter release (Denoyer et al., 2020), or a gradual shift in synaptic weights due to cumulative shifts in the timing of spikes (Modolo et al., 2013). We argue that the lasting effects of tACS represent an extraordinary window of opportunity: leveraging and optimizing such lasting effects would greatly impact future uses of tACS as a treatment. It would be possible indeed to envision tACS therapy optimized to induce maximally lasting effects in the minimal number of stimulation sessions possible, which would certainly increase the feasibility and compliance of at-home use.

Among those mechanisms, one critical point is that it has recently been pointed out that the reported effects of tACS could be explained not by central (i.e. modulation of neuronal activity) effects of the induced electric fields, but instead by the direct activation of peripheral (i.e. cranial) nerves, thereby indirectly producing effects on neuronal excitability. This provocative explanation has gained significant attention, since this could jeopardize efforts in the tACS field, such as those regarding the optimization of electrode montages to ends of personalizing therapy. The rationale for this proposal is the following: tCS induces, at the scalp level, electric fields that can reach levels on the order of 20 V/m, which can activate peripheral (cranial and cervical) nerves. In turn, peripheral nerve activation can cause the release of noradrenalin by the locus coeruleus, which then induces a modulation of excitability and plasticity in the neocortex (van Boekholdt et al., 2021). Therefore, if this scenario is correct, peripheral nerve activation could result in comparable neurophysiological effects in terms of excitability modulation and cognitive performance, while completely bypassing the direct modulation of neuronal membranes and entrainment of oscillations, i.e. a whole different chain of neurophysiological processes. However, as discussed below, the current state of the literature is that tCS exerts its effects, at least in part, through a direct modulation of neuronal activity as evidenced by effects on spike timing in monkeys, or by alpha oscillations in ferrets (Huang et al., 2021; Johnson et al., 2020). Let us note that this does not exclude a possible contribution from the aforementioned activation of cranial nerves.

### *Moving tCS forward: experimentally testable hypotheses*

As aforementioned, recent experimental evidence supports that tACS exerts its effects through a central mechanism, without fully excluding a role for peripheral nerve. In order to provide a definite view on this problem, we propose the following solution: setting up an experiment in which participants would be performing the exact same cognitive performance task using a different brain stimulation modality: either tACS (which is electric) or a recently proposed technique called transcranial alternating magnetic stimulation (tAMS, which is magnetic) that avoids the issue of inducing any sensation at the scalp level and does not use any electrode but a coil instead (Legros, 2023). Ideally, the simultaneous use of high-resolution electroencephalography (HR-EEG) might be an asset, providing altogether a neurophysiological detailed characterization of the modulation of cortical oscillations along with task performance. The magnetic field flux density of tAMS should be calibrated so that the *in situ* electric field induced matches the tACS electric field norm at the cortical level. Comparable changes in HR-EEG-derived measures of neuronal activity and associated cognitive performance would provide a powerful confirmation of the "central effects" hypothesis for tACS.

Another word of caution regarding the reported effects of tACS on human neurophysiology and cognitive performance, and the associated underlying mechanisms, is the potential of phosphene perception to act as a confound in a significant number of studies, since the sensitivity of the retina to low induced electric fields (which is precisely the reason why phosphene perception is used as a basis for the safety guidelines regarding low-frequency electromagnetic fields exposure, as discussed extensively in (Legros, 2023)) triggers visual perceptions that are associated with changes in electrophysiological signals and possible interfere with behavior during tasks. Therefore, there is a possibility that a number of the aforementioned studies in monkeys and humans might suffer from this bias. Let us mention that some studies have shown special care to avoid retina-related effects (Fiene et al., 2020).

New tACS design may tap further into dynamical systems mechanisms of coupled oscillations that are established in related domains in behavioral neuroscience and biology in general. It may prompt novel approaches to engaging endogenous brain dynamics using tACS involving multiple phases and frequencies (Riddle & Frohlich, 2021). Building on existing observations of Arnold Tongue in tACS entrainment of neuronal activity (Ali et al., 2013; Huang et al., 2021), a natural next step is to test the empirical relevance of Arnold Tongues in a multifrequency (polyrhythmic) framework. It is a signature of biological systems, especially the brain, to exhibit multiple coordinated rhythms (Buzsáki, 2006; Glass, 2001). Arnold Tongue structures have primarily been used as a theoretical basis for understanding polyrhythmic biological coordination, i.e., two oscillatory processes exhibit mode-locking at integer frequency ratios such as 1:1, 1:2, 2:3 etc. The most stable mode is 1:1, corresponding to the thickest Arnold Tongue, which is the focus of existing tACS studies. The relative stability of different frequency ratios forms a hierarchy, namely the Farey tree, where lower order ratios exhibit greater stability (wider Arnold Tongues). Empirically, mode-locking has been shown as a "devil's staircase" – the observed ratio between the entrained and the forcing oscillator plateaus at preferred ratio as the frequency of the forcing oscillator increases continuously (Assisi et al., 2005; de Guzman & Kelso, 1991; Peper et al., 1995); multiple modes can coexist in larger groups of oscillators with diverse natural frequencies (Zhang et al., 2018). A continuous frequency sweep using tACS may reveal higher-order Arnold Tongue structures, which may in turn create the opportunity to use tACS at a single frequency to modulate multiple cross-frequency relationships.

Another cross-frequency mechanism, more extensively observed in cognitive neuroscience, is phase-amplitude coupling (PAC) (Jensen & Colgin, 2007; Tort et al., 2010). That is, the amplitude of a fast oscillation is modulated by the phase of a slower oscillation. Slower oscillations were thought to facilitate attentional selection of sensory or lower-level activities (Schroeder & Lakatos, 2009). However, causal manipulation of PAC was scarcely employed. Recently, Riddle et al. (2020) showed that cross-frequency tACS (CF-tACS) enhanced targeted delta-beta and theta-gamma phase-amplitude coupling, which modulated distinct components of behavioral performance in a working memory task. This work demonstrates the potential of using CF-tACS to enhance the specificity of target engagement compared to single-frequency tACS. The underlying neurophysiological mechanisms of CF-tACS remains to be explored. Computational models and animal experiments could help demonstrate the effect of CF-tACS on neuronal dynamics.

The distinction between mode-locking in the Arnold Tongue framework and PAC, as mechanisms of cross-frequency coupling, is that the former requires phase-coupling while the latter does not. Phase-coupling is most often used to describe generalized synchronization between oscillators at the same frequency. In cross-frequency phase-coupling, the faster oscillation is coupled to multiple phases of the slower oscillation. PAC does not require phase-coupling. In observational studies, that is, without causal manipulation of the oscillation itself, it is difficult to distinguish between genuine cross-frequency coupling and non-sinusoidal waveforms from a purely statistical point of view (Scheffer-Teixeira & Tort, 2016). tACS with continuous parameter manipulations would be invaluable for distinguishing cross-frequency interaction from artifacts, and amplitude-coupling from phase-coupling. In particular, one may expect the statistical measure of cross-frequency coupling to make discrete jumps rather than smoothly varying with continuous change of tACS parameters – a key experimental approach to detecting genuine coupling in nonlinear dynamical systems (Kelso, 1995).

Beyond the classical theories of synchronization and mode-locking, another cross-frequency mechanism of interest is metastability (Kelso, 2012; Tognoli & Kelso, 2014). Brain oscillations are not sustained, but rather, intermittent (Jones, 2016). Metastability is a dynamical systems explanation for intermittent synchronization, which may manifest in part as intermittently emerging macroscopic oscillations across scales (Tognoli & Kelso, 2009; Tognoli & Kelso, 2014; Tognoli et al., 2018; Zhang et al., 2019; Zhang et al., 2020). The regime of metastability falls right outside the boundary of Arnold Tongues: the natural frequency of two oscillators must be different and the coupling between them need to be sufficiently weak. The duration of epochs of (apparent) synchronization or macroscopic oscillations is inversely related to the square-root of the distance to the nearest Arnold Tongue (a feature of saddle node bifurcations). Here again, parametric manipulation of tACS waveform presents unique opportunities for examining such cross-frequency mechanisms. One study examined the coupling of two rhythmic network with long-range connections with biophysically plausible delays (Kutchko & Fröhlich, 2013). The two networks exhibit transitions between different states; tACS applied in phase to both networks modulated the state transition probabilities. This suggests that tACS may switch interconnected networks between different states.

Finally, we present in Figure 2 an illustration of possible modeling efforts that could contribute to clarify the acute effects of tACS, at the single-cell level, while being experimentally testable. The use of model-guided techniques has been increasingly adopted by the tACS community, as an effort to bridge the reported effects in humans with corresponding neurophysiological mechanisms. Such modeling endeavors include models based on the classical, robust formalism by Hodgkin and Huxley (Hodgkin & Huxley, 1952) have already initiated investigating how neuronal activity, as quantified using various metrics, was modulated by low-intensity electric fields such as those induced by tACS (Tran et al., 2022). We propose that an exhaustive, quantitative characterization of phase entrainment as a function of tACS frequency should be performed in all the main cellular types found in the neocortex to establish a sort of "dictionary" of the most efficient tACS frequencies to use with the intent to achieve maximal phase entrainment in targeted neuronal sub-populations.

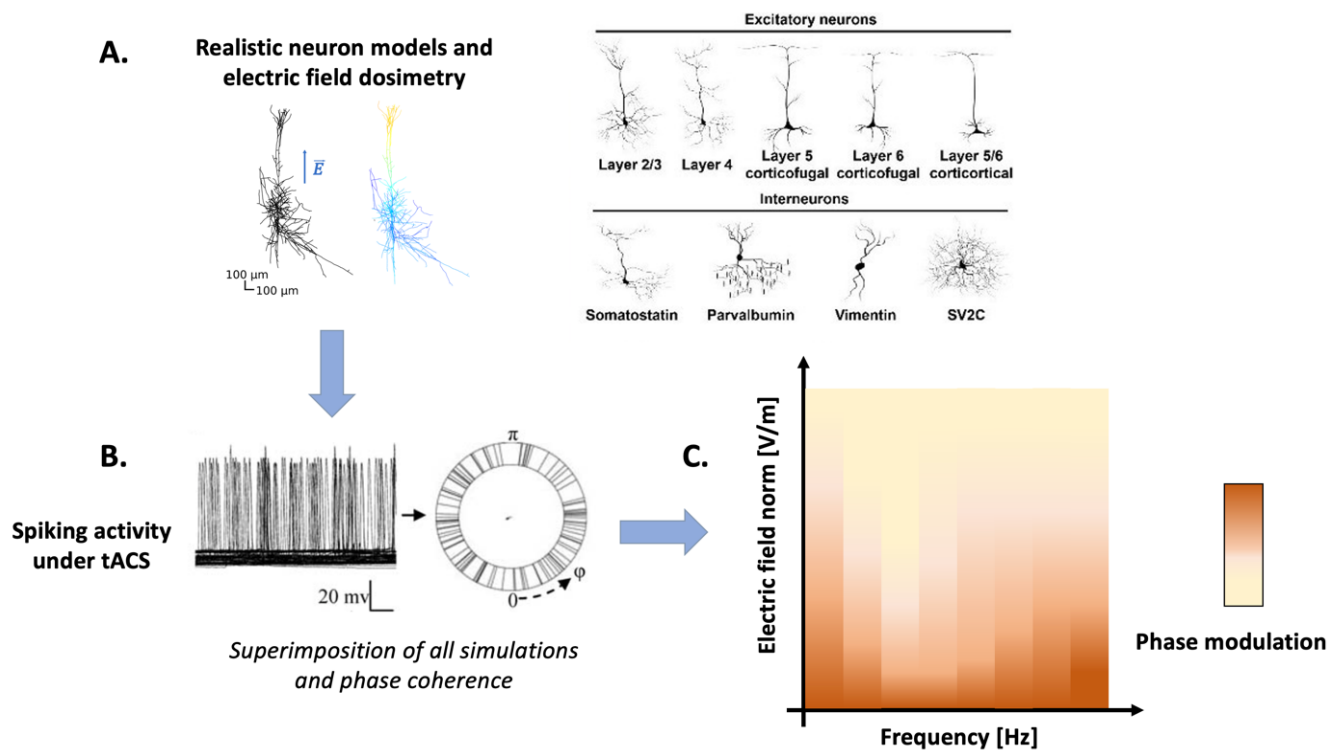

**Figure 2.** Proposed of an experimentally testable, model-guided approach to elucidate the cellular-scale basis of tACS acute effects.

**Discussion and concluding remarks**

Thanks to considerable research efforts in the field of tACS, tackling the issue of mechanisms using *in silico*, *in vitro*, *in vivo* and clinical approaches, the technology and understanding of its effects have significantly matured, some neurophysiological and behavioral effects can be reliably replicated, while some new challenges emerged. From the results discussed in this review, it can be claimed with a confidence that tACS can modulate the endogenous activity of specific brain circuits and associated behavior, pending that the stimulation frequency is close to the said endogenous activity. Also, a single session of tACS can induce lasting effects up to approximately one hour, which is a replicated result and can also be confidently claimed. Notably, a significant evolution in the field which has likely increased reproducibility is the use of detailed dosimetric models that provide a spatial estimation of the induced electric field in the main brain tissues, and taking into account the most exquisite details of cortical morphology. Those dosimetric models provide an invaluable tool that provides an accurate quantification of the performance of specific electrode montages in focusing the field in specific targeted regions. Combined with the neurophysiological knowledge regarding the endogenous frequencies of the targeted region, tACS can target brain circuits in space and time to maximize the probability to obtain pre-determined effects. One word of caution is that, if peripheral effects were identified as being a major driver in the reported effects of tACS, it would question the need for such for sophisticated, patient-specific electric field modeling. However, from the available evidence at the time of this review, this possibility appears unlikely, even if we cannot exclude that effects from peripheral nerve activation explain, at least in part, some of the reported effects of tACS. Providing a definite answer regarding this issue will require, for example, the use of non-invasive brain stimulation technologies that have the potential to modulate ongoing oscillations without inducing any noticeable perception at the skin level (e.g., using sinusoidal magnetic fields delivered by Helmoltz-like coils, see (Legros, 2023)).

Interestingly, the tACS mechanisms that have been identified have been confirmed at several spatial scales (from single neurons to local oscillations involving millions of neurons) using a variety of experimental techniques, and also in different species (mouse, rat, ferret, monkey). This convergence provides a support for those effects. Notably, the most established tACS neurophysiological effects are all *acute* effects, i.e. immediate effects (e.g., effects on phase coherence, spike timing); while the most promising (and least understood) results will undoubtedly come from a thorough understanding of *lasting* effects. While there is partial evidence for effects on synaptic plasticity modulation, a complete and causal mechanistic chain of mechanisms explaining lasting effects is yet to discover, and will require extensive experiments at various spatial scales (unit activity, LFP) probably using a combination of techniques (patch-clamp, optogenetics, application of pharmacological agents). From the aforementioned issues, carefully-designed protocols are also required to exclude the contribution of peripheral nerve activation, and to confirm that tCS can modulate precisely the oscillations in target brain circuits underlying specific functions (Riddle and Frohlich, Brain Res 2021). We argue that the issue of lasting effects induced by tACS is the most crucial and pressing one, since this would pave the way for protocols optimizing the duration of such after-effects, thereby minimizing the application duration of tACS and increasing its potential clinical, or even home use (Jones et al., 2022; Palm et al., 2018).

In conclusion, tACS is a promising technique, in that it is not only affordable due to the relative low technical complexity, but also an safe technique that could have a tremendous impact in clinical practice, if it indeed mainly acts by the modulation of cortical circuits directly, and that its mechanisms can be understood to enable technical and practical optimization of protocols. In addition, while significant technological advances have been in the field, such as the possibility to perform "high-resolution" stimulation with patient-specific montages, there is still room for further innovations, that could for example aim at increasing the stimulation intensity while remaining tolerable for subjects by using next-generation montages or stimulation waveforms. Finally, improved temporal targeting through closed-loop tACS has very recently emerged as a promising treatment in psychiatry (Schwippel et al., 2024).

**Acknowledgments**

This study is supported by the Labex Cominlabs project "PKSTIM" and "SP-STIM". FF and MZ report funding support from NIMH R01MH124387